\documentstyle[aps,epsf]{revtex}
\draft

\begin{document}
\title{Nonadiabatic Transition Probabilities in the Presence of 
\protect\\ Strong Dissipation at an Avoided Level Crossing Point}
\author{Keiji Saito}
\address{Department of Applied Physics, School of Engineering \\ 
University of Tokyo, Bunkyo-ku, Tokyo 113-8656, Japan}
\author{Yosuke Kayanuma}
\address{Department of Mathematical Science, Graduate School of Engineering \\
Osaka Prefecture University, Sakai 599-8531, Japan}
\maketitle
\begin{abstract}
Dissipative effects on the nonadiabatic transition for the two and three 
level systems are studied. When the system is affected by a strong 
dissipation through the diabatic states, the exact transition probability 
is enumerated making use of the effective master equation.
In the two-level system, we consider the case where the external 
field is swept from not only a negative large value but also from the resonant field, 
and the exact transition probabilities in these cases are derived. 
The transition probabilities are derived for the three-level system where 
the three diabatic states form only one avoided level crossing point. 
These probabilities are compared with the one in the pure quantum case 
obtained by Carroll and Hioe.
\end{abstract}
\vspace*{0.5cm}
\section{Introduction}
Nonadiabatic transition at an avoided level crossing point plays a 
crucial role in quantum dynamical changes of states, and yields the variety of
phenomena in physics and chemistry.  
The well-known Landau and Zener (LZ) transition probability 
clarifies the roles of the energy gap and the sweeping velocity of the external field in the 
nonadiabatic transition of two-level system \cite{Landau,Zener,St,ZN}.
Although the LZ transition probability is given in the two-level system, 
it is approximately applicable to multi-level system where 
the avoided level crossings are effectively well described by only
localized two levels.
Hence it is adopted in the analyses of many experiments which treat time dependent phenomena,
such as collision of particles \cite{AP83,C74}, optics \cite{RKLK81,B_etal95}, and
magnetic phenomena \cite{mn12,fe8,miya9596,RMSGG97,SMD00}. 
For general multi-level systems where many levels can simultaneously
affect on each other, different formulas of transition probabilities are derived for several models, i.e., 
the one where only one level interacts with a band of levels \cite{D66,O66,KF85}, 
the generalized model for this model \cite{DO95,DKO95} and the bow-tie model where many 
levels form only one avoided level crossing \cite{CH86,ON97}. We should 
also note Brundbler and Elser's hypothesis which states that the survival probability 
of the diabatic state with maximum or minimum slope is described by the
exponential form determined 
by only the velocity and the off-diagonal elements in the Hamiltonian \cite{BE93}. 

On the other hand, we must also consider the effect of dissipation,
since real experiments are always exposed to thermal environment.
The thermal environment causes decoherence and the inevitable deviation of
transition probability from the one of pure quantum case occurs. 
This modification becomes significant in real experiments such as
the adiabatic rapid passage with phonon couplings\cite{ARP}, the nonadiabatic transitions in localized centers in solids\cite{K82}, and nonadiabatic magnetization
process in molecular magnets such as Mn$_{12}$ and Fe$_{8}$ \cite{W00,N01,CH,BR}. 
Kayanuma studied such thermal noise effect for the two-level Landau-Zener model, 
and derived a formula for the effective transition probability in the limiting case of
strong damping dissipation by perturbative approach\cite{K84}. 
The effective transition probability becomes $1/2$ in the adiabatic
limit due to the dissipation effect, whereas it converges to the
asymptotic expression of the LZ probability in the fast sweeping limit.
Ao and Rammer carried out first principle calculation to investigate
temperature dependence of the transition probability of the two-level
system with phonon reservoir which corresponds to the Kayanuma's 
situation in the high temperature limit in case of the Ohmic spectral density. 
Especially they found some compensation effect 
that the transition probability for zero temperature becomes 
the same value as the Landau-Zener probability \cite{AR91}.

In this paper, we study such thermal noise effect in not only the two-level system but also the three-level system. 
Thereby we try to investigate the effect of multi-level with thermal noise. 
We exploit the method to analyze strong dissipation effect using an effective master equation 
instead of the perturbation approach adopted in previous studies \cite{K82,K84,AR91}. 
We show that the effective master equation approach is very convenient for deriving the transition probability 
in the strong damping limit. Using this approach 
we first reproduce Kayanuma's formula in the two-level system 
when the external field is reversed from large negative value to large positive value. We next consider 
the situation where the field is swept from resonant field (zero field) 
to large positive field, and derive
exact transition probability. As a result the exact relations between these cases are found. 
The three-level model we consider is the same model 
as Carroll and Hioe considered \cite{CH86}. In the model, three diabatic
states form only one avoided level crossing point. Therefore the
transition mechanism is quite different from the LZ mechanism which describes transitions between local two levels. 
Therefore we see the effect of multi-level not only in pure 
quantum case but in dissipative case.
We adopt the effective master equation 
approach and derive the transition probabilities in the strong damping limit. The probabilities are
compared with the one of pure quantum case obtained by Carroll and Hioe.
The expression of the probabilities 
are always the same regardless of level structures, 
although in pure quantum case the expressions of the transition probabilities 
show some variations.

This paper is organized in the following way. In section II, we derive
the master equation when strong noise couples with the system and 
compare it with the master equation for the system with phonon
reservoir.  Section III is devoted to the problem for two-level system,
and we derive the transition probabilities for three-level system in
section IV. Summary and brief discussion are given in section V.  
\section{Master Equation}
We derive the master equation for the system with dissipation. 
Throughout this paper, we study the transition probability by solving the
master equation. The master equation we shall consider is derived 
for various types of dissipative environments such as stochastic noise
field and
phonon reservoir. We here choose a stochastic noise field as a source of
dissipation and rigorously derive the master equation.  
As shown in Appendix A, this master equation can be obtained in case of phonon reservoir 
with the Ohmic type spectral density at very high temperatures. 
The correlation of the stochastic noise is assumed to be very short. This situation is described as
\begin{eqnarray}
{\cal H}_{\rm tot} &=& {\cal H} (t) + \sum_{\ell }\xi_{\ell} (t) X_{\ell} ,
\label{htot} \\
\langle \xi_{\ell} (t) \xi_{m} (t' ) \rangle &=& 
2\gamma_{\ell}  \delta_{\ell , m}\, \delta (t- t' ) 
\label{noise},
\end{eqnarray}
where $\xi_{\ell} (t)$ is a noise which affects on the system through the 
$\ell$th operator $X_{\ell} $. The matrix $X_{\ell}$ is diagonal in the diabatic bases of the Hamiltonian ${\cal H} (t)$ 
so that the computability $\left[X_{\ell} , X_{\ell '} \right] = 0$ 
is satisfied for arbitrary $\ell$ and $\ell '$. 
The noise is supposed to be the white Gaussian process.
We start with the Von-Neumann equation for the density matrix in the
interaction picture ($\hbar=1$ here and hereafter),
\begin{eqnarray}
\frac{\partial \rho^{\rm (I)} (t)}{\partial t} 
&=& \sum_{\ell}\xi_{\ell} (t) {\cal L}_{\ell} (t ) 
\rho^{\rm (I) } \label{feq}\\ 
\xi_{\ell} (t) {\cal L}_{\ell} (t)\rho^{\rm (I) }
&=& -i \xi_{\ell} (t) \left[ \rho^{\rm (I)} (t) , X_{\ell} (t) \right], \\
 \rho^{\rm (I)} (t) &=& \exp_{\leftarrow} 
\left( -i \int_{t_0}^{t} \, du \, {\cal H} (u) \right)
\rho (t)\exp_{\rightarrow} 
\left( i  \int_{t_0}^{t} \, du \, {\cal H} (u) \right) , \\
 X_{\ell}(t) &=&  \exp_{\leftarrow} 
\left( -i \int_{t_0}^{t} \, du \, {\cal H} (u) \right)
X_{\ell} \exp_{\rightarrow} 
\left( i  \int_{t_0}^{t} \, du \, {\cal H} (u) \right)
\end{eqnarray}
Here $\exp_{\leftarrow} $ and $\exp_{\rightarrow}$ express the time
ordered product of exponentials.
In the case of white Gaussian process (\ref{noise}), 
there are several approaches to derive master equation \cite{K63,N64}. 
Here we use the Novikov's relation,
which holds for arbitrary function $g ( [\xi ] , t)$ \cite{N64},
\begin{eqnarray}
\langle \xi g ([\xi ] , t) \rangle &=& \int_{t_0}^{t} \, d t'  
\langle \xi (t) \xi (t') \rangle\langle \frac{\delta \, g ([\xi] , t)}
{\delta \,\xi (t')  }\rangle  ,
\end{eqnarray} 
where the symbol $[\xi ]$ means that $g ([\xi ] , t)$ is 
a function of the process of noise
$\xi$, and $\langle ... \rangle$ means the average over the noise $\xi(t)$. 
By use of this mathematical formula, the average over
noise for Eq.(\ref{feq}) is reduced to,
\begin{eqnarray}
\frac{\partial\langle  \rho^{\rm (I) } (t)\rangle  }{\partial t}
&=& \sum_{\ell}\langle \xi_{\ell} (t) {\cal L}_{\ell} (t ) 
\rho^{\rm (I) } (t) \rangle 
=\sum_{\ell}\langle \xi_{\ell} (t) {\cal L}_{\ell} (t ) \, 
\exp_{\leftarrow}\left( \sum_{\ell '}\int_{t_0}^{t}\, du \, \xi_{\ell '} (u) 
{\cal L}_{\ell ' } (u ) \right) \, \rho^{\rm (I) } (t) \rangle 
\nonumber  \\
&=&\sum_{\ell}{\cal L}_{\ell} (t ) \int_{t_0}^{t} \,  
d t'  \langle \xi_{\ell} (t) \xi_{\ell} (t') \rangle
\langle \exp_{\leftarrow}\left( \sum_{\ell '}\int_{t'}^{t}\, du \, \xi_{\ell '} (u) 
{\cal L}_{\ell ' } (u ) \right) {\cal L}_{\ell} (t' ) \rho^{\rm (I)} (t' )
   \rangle  
\nonumber \\
&=& \sum_{\ell}\gamma_{\ell} 
{\cal L}_{\ell}^{2} (t ) \langle \rho^{\rm (I)} (t )  \rangle .
\end{eqnarray}
Here we used Novikov's relation and the properties of the white noise. 
Thus we arrived at the master equation in the Shr\"{o}dinger picture of the density matrix 
\begin{eqnarray}
\frac{\partial \rho (t)}{\partial t}
&=& -i\left[ {\cal H }(t) , \rho(t)\right]
-\sum_{\ell}
\gamma_{\ell} \left[ X_{\ell}, \left[ X_{\ell}, \rho(t) \right]\right]
\label{maseq} , 
\end{eqnarray}
in which we denoted $\rho (t)$ for $\langle \rho (t) \rangle $, omitting 
the symbol of average over noise $\langle \quad \rangle $.
In the following sections, we focus on the noise 
with short correlation (\ref{noise}) and large amplitude of $\gamma $,
namely the strong damping (SD) limit. 
This is equivalent to the phonon reservoir with high temperature.
We investigate the properties of the nonadiabatic transitions 
based on the master equation 
(\ref{maseq}).

\section{Dissipative Effects in Two-Level System}
\subsection{Transition Probability}
In this section, we consider the noise effect in the two-level system.
For two-level system, the transition
probabilities in the strong damping case have been 
studied by several authors using perturbation approach \cite{K82,AR91}.
Here we alternatively adopt the different approach, deriving
the effective master equation which simplifies derivations 
of transition probabilities for the case. 
Here two situations are studied. We first 
consider the familiar situation that the external field 
is reversed from $-\infty$ to $\infty$. In this case, 
we demonstrate that Kayanuma's transition 
probability $P_{\rm SD}^{-+}$ is reproduced easily using the effective master 
equation.
We second consider somewhat unfamiliar but realizable situation that 
the external field is swept from $0$ to $\infty$ when 
initially a diabatic state is occupied. In magnetic
system, this situation corresponds to the case where 
a spin is saturated to the down (or up) state under very strong field 
and next the field is switched off and the field is
swept linearly in time from zero field.

The model we shall consider in this section is described as,
\begin{eqnarray}
{\cal H}_{\rm tot} (t) &=& {\cal H}(t) + \xi (t) \sigma^{z}, \\
{\cal H}(t) &=& -vt \frac{1}{2}\sigma^{z} + \Gamma \frac{1}{2}\sigma^{x} , 
\end{eqnarray}
where $\sigma^{\alpha}$ is the $\alpha (=x,y,z)$ component of the Pauli
matrix. The diabatic states correspond to the down state $|1\rangle $ 
and up state $|2 \rangle$ which satisfy $\sigma^{z}|1\rangle = -
|1\rangle $ and $\sigma^{z}|2\rangle = |2\rangle $,
respectively. $\Gamma$ is the transverse field which is responsible for
the tunneling between the diabatic states.  
Here we take only $\sigma^z $ as the operator on which the noise acts in
(\ref{noise}). 
Thus the master equation (\ref{maseq}) is concretely written as, 
\begin{eqnarray}
\frac{\partial }{\partial t} \rho (t) 
&=& -i \frac{1}{2} \left[ vt \sigma_z + \Gamma \sigma_x , \rho (t)
\right] -\frac{\gamma}{2}
\left( \rho (t) - \sigma_z \rho (t) \sigma_z \right) \label{master} .
\end{eqnarray}
We define the following variables:
\begin{eqnarray}
c_1 &=& \rho_{11}-\rho_{22} , \\
c_2 &=& \rho_{12} , \\
c_3 &=& \rho_{21} . 
\end{eqnarray}
The time evolutions of these variables are determined by 
the differential equations:
\begin{eqnarray}
\dot{c}_{1} &=& -i\Gamma (c_3 -c_2 ) , \label{c1}\\
\dot{c}_{2} &=& (-ivt - \gamma) c_2  + \frac{i\Gamma}{2} c_1 , \label{c2}\\
\dot{c}_{3} &=& (ivt - \gamma) c_3  - \frac{i\Gamma}{2} c_1 \label{c3} . 
\end{eqnarray}
Here we consider the SD limit, $\gamma \rightarrow \infty$.
We first consider the variable ${c}_{2} (t) $, which 
is formally solved from Eq.(\ref{c2}). We can approximate 
it by partial integral; 
\begin{eqnarray}
{c}_{2} (t) &=& c_{2} (t_0 ) +\frac{i\Gamma}{2}
e^{-ivt^2 /2 - \gamma t} \int_{t_0}^{t} \, du \, 
e^{iv u^2 /2 + \gamma u}  c_1 (u)  \nonumber \\
&=& c_{2} (t_0 ) +\frac{i\Gamma}{2} e^{-ivt^2 /2 - \gamma t} \left(
\left[e^{iv u^2 /2 + \gamma u}\frac{c_1 (u)}{ ivu + \gamma}  \right]_{t_0}^{t} 
- \frac{\Gamma}{2v}\int_{t_0}^{t}\, du \, 
e^{iv u^2 /2 + \gamma u  }
\frac{d}{du} \left(\frac{c_1 (u)}{ivu + \gamma }  \right) \, \right)
\nonumber \\
&\sim & c_{2} (t_0 ) +\frac{i\Gamma}{2}\left(
\frac{c_1 (t)}{ ivt + \gamma} - \frac{\frac{d}{dt} \left(\frac{c_1 (t)}{ivt + \gamma }  \right)} {ivt + \gamma  }  + \cdots
\right)
\nonumber \\
&\sim& c_{2} (t_0 ) + \frac{i\Gamma}{2}  \frac{c_1 (t)}{ivt + \gamma } .
\end{eqnarray}
Here we used the fact that the term $e^{- \gamma (t - t_0 )}$ is negligible due to
large $\gamma$, and we neglected the higher order terms of $(ivt + \gamma
)^{-1}$ \cite{note}. 
When the diabatic state $|1\rangle $ is initially occupied, namely,
\begin{eqnarray}
 c_1 (t_0 ) = 1 , \qquad  c_2 (t_0 ) = c_3 (t_0 ) = 0, 
\end{eqnarray}  
$c_2 (t)$ and $c_3(t)$ are approximated as,
\begin{eqnarray}
{c}_{2} (t) &=& \frac{i\Gamma}{2}  \frac{c_1 (t)}{ivt + \gamma } , \label{r_c2}\\
{c}_{3} (t) &=& \frac{-i\Gamma}{2}  \frac{c_1 (t)}{-ivt + \gamma } \label{r_c3}.
\end{eqnarray}
These relations lead us to the simplified equation for
the diabatic states. By substituting the relations (\ref{r_c2}) and
(\ref{r_c3}) into the equation (\ref{c1}), we arrive at the effective
master equation :  
\begin{eqnarray}
\dot{c}_1 (t) &=& \frac{\Gamma^2}{2iv} \left\{ 
\frac{1}{t+ i \gamma /v} - \frac{1}{t - i \gamma / v}
\right\} c_1 (t)  \label{eme}.
\end{eqnarray}
Now let us consider the first problem, i.e. $t_0 = - \infty$.
In this case, we can readily integrate the master equation (\ref{eme}) 
to get, 
\begin{eqnarray}
c_1 (\infty ) = \exp \left( - \frac{\pi \Gamma^2 }{v}\right) .
\end{eqnarray}
We now consider the tunneling probability from the state $|1 \rangle$
at $t=-\infty$ to the state $|2 \rangle $ at $t=\infty$. 
This corresponds to the value of 
$\rho_{22} ( \infty)$. 
By using the conservation of probability ${\rm Tr} \rho = 1$, 
this transition probability $P_{\rm SD}^{-+} ( \equiv \rho_{22} ( \infty))$ is obtained,
\begin{eqnarray}
P_{\rm SD}^{-+} &=& \frac{1}{2} \left(  1 - \exp \left( - \frac{\pi \Gamma^2 }{v}\right) 
\right) . \label{psd-+}
\end{eqnarray}
This is nothing but Kayanuma's transition probability \cite{K84}

For the second case where the field is swept from resonant field, 
i.e., $t_0 = 0$, $c_1 (\infty )$ is readily calculated 
as 
\begin{eqnarray}
c_1 (\infty ) = \exp \left( - \frac{\pi \Gamma^2 }{2v}\right) ,
\end{eqnarray}
which gives the transition probability $P_{\rm SD}^{0+}$,
\begin{eqnarray}
P_{\rm SD}^{0+} &=& \frac{1}{2} 
\left(  1 - \exp \left( - \frac{\pi \Gamma^2 }{2v}\right) \right) .
\label{psd0+}
\end{eqnarray}
The validity of these probabilities (\ref{psd-+}) and (\ref{psd0+}) is confirmed by numerically integrating the master 
equation (\ref{master}). In figure 1, we present numerical data and
the analytical results given in (\ref{psd-+}) and (\ref{psd0+}) as a function of sweeping velocity. Here $\Gamma $ is $0.01$,
and $\gamma$ is $10.0$. We show not only the
dissipative case but also pure quantum case ($P_{\rm LZ}^{-+}$ and
$P_{\rm LZ}^{0+}$). We see that formulas (\ref{psd-+}) and (\ref{psd0+}) agree with the numerical results almost perfectly. 
\begin{figure}
\begin{center}
\noindent
\epsfxsize=9.0cm \epsfysize=6.0cm
\epsfbox{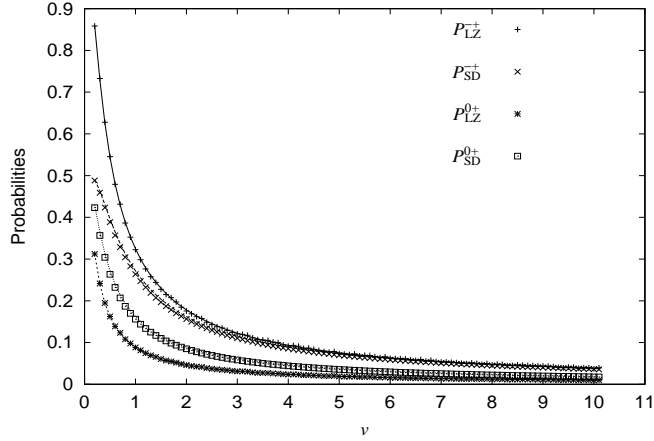}
\end{center}
\caption{Transition probabilities as  a function of sweeping velocity $v$.
 Parameters  are taken as $\Gamma = 0.01$, and  $\gamma = 10.0$. Points are numerical 
 data, and lines are the analytical values. Here
the subscript 'LZ' means that it is pure quantum case. $P_{\rm LZ}^{0+}$
is the probability in the case where the external field is swept from
zero field in pure quantum case. }
\end{figure}

The variable $c_1 (t)$ is directly connected with the magnetization 
$M(t) = {\rm Tr \, \sigma^{z} \rho (t)}$. 
We obtain the magnetization process solving (\ref{eme}) as,
\begin{eqnarray}
M^{-+}(t) &=& -\exp\left[ \frac{\Gamma^2 }{v}\arctan \left(\frac{\gamma}{vt} \right) \right] , \\
M^{0+}(t) &=& -\exp\left[ \frac{\Gamma^2 }{v}
\arctan \left(\frac{\gamma}{vt} \right)  - \frac{\pi}{2}\right] ,
\end{eqnarray}
where the function $y = \arctan x$ is defined in the region of 
$x \in [-\infty , \infty]$ and $y \in [-\pi , 0 ]$.
This shows that the the magnetization process depends on the noise
strength $\gamma$, whereas final magnetization does not. 
This is also numerically confirmed.

\subsection{Relation between $P_{\rm SD}$ and $P_{\rm LZ}$}
The exact relation of the transition probabilities for pure
quantum case and in strong damping case is discussed.
When the field is reversed from large negative value 
without dissipation, the transition probability is denoted by
$P_{\rm LZ}^{-+}$. Here the subscript 'LZ' means that it is the 
pure quantum case. In the case of
the field swept from the resonant point without dissipation,
the transition probability is written as $P_{\rm LZ}^{0+}$. 
As briefly shown in Appendix B, 
the transition probability $P_{\rm LZ}^{0+}$ is calculated as,
\begin{eqnarray}
P_{\rm LZ}^{0+} &=& \frac{1}{2}\left( 1 - 
\exp\left( -\frac{\pi\Gamma^2}{4v}\right)\right) ,
\end{eqnarray}
Thus transition probability in each case is written as follows,
\begin{eqnarray}
\begin{array}{ccccc}
P_{\rm SD}^{-+} = \frac{1}{2}
\left( 1 - \exp\left( -\frac{\pi\Gamma^2}{v}\right) \right)
&\quad & 
P_{\rm LZ}^{-+} &=& 
1 - \exp\left( -\frac{\pi\Gamma^2}{2v}\right)  \\
P_{\rm SD}^{0+} = \frac{1}{2}
\left( 1 - \exp\left( -\frac{\pi\Gamma^2}{2v}\right) \right)
&\quad & 
P_{\rm LZ}^{0+} &=& \frac{1}{2}
\left( 1 - \exp\left( -\frac{\pi\Gamma^2}{4v}\right) \right)  \label{lz0+}.
\end{array}
\end{eqnarray}
These asymptotically exact probabilities satisfy the following relations (see also the Fig.1),
\begin{eqnarray}
P_{\rm SD}^{-+} &\le& P_{\rm LZ}^{-+} \label{enq-+} \label{ene1} , \\
P_{\rm SD}^{0+} &\ge& P_{\rm LZ}^{0+} \label{enq0+} \label{ene2} .
\end{eqnarray}
The inequality (\ref{ene1}) means that the dissipation reduces
tunneling that the state remains in the ground state. 
When $v\ll \Gamma^2$ in the pure quantum case, 
almost adiabatic tunneling from $|1\rangle$ to $|2\rangle$ occurs,
whereas in the presence of dissipation,
thermal excitation from ground state to the excited state represses 
such adiabatic evolution. Thus the inequality (\ref{ene1}) is easily
understood. On the other hand, the inequality (\ref{ene2}) ibdicates  
the opposite property. 
This effect is intuitively explained as follows. 
The initial state $\psi (0)\, (= | 1 \rangle)$ is the superposition between 
the ground state $| G (0) \rangle$ and the excited state $| E (0)\rangle$ ,
\begin{eqnarray}
\psi (0) = \frac{1}{2} \left(|G(0)\rangle + |E(0) \rangle \right) \label {initpsi}
\end{eqnarray} 
In case of almost adiabatic evolution in the pure quantum case, $v \ll 1$, 
the state of the system almost follows
such superposition at $t$,
\begin{eqnarray}
\psi (t) \sim  \frac{1}{2} \left( e^{i\phi_1 (t) }|G(t)\rangle + 
e^{i\phi_2 (t) }
|E(t) \rangle \right).
\end{eqnarray}
where $e^{i\phi_1 (t)} $  and $e^{i\phi_2 (t) }$ are the dynamical phases.
Since $|G(t)\rangle \rightarrow |2 \rangle $,
$|E(t)\rangle \rightarrow |1 \rangle $ in the limit of $t\rightarrow \infty$,
the maximum value of transition probability is $\frac{1}{2}$ 
in the pure quantum evolution. 
In the presence of the dissipation, the noise also induces such 
uniform distribution, because the dissipation we now consider can be 
regarded as thermal effects with very high temperature.
As a result the tunneling probability is larger in presence of the dissipation.
We may say that the inequality (\ref{ene2}) is a consequence of the 
special initial state, because if the initial state is not a diabatic
state, e.g. $\psi (0) = |G(0)\rangle$, 
the inequality (\ref{ene2}) is not realized. 
We expect these characteristic relations (\ref{ene1}) and (\ref{ene2}) 
will be verified in real experiments.
\section{Three-Level System}
We here apply our approach to the multi-level systems, and investigate 
the dissipative effect on many levels. For the sake of definiteness, we focus our attention on the properties of 
the three-level system, where all diabatic levels form only one avoided level
crossing point. Because the avoided crossing structure is not formed by
localized two states, the transition mechanism is very different from
the LZ mechanism. Therefore this three-level system will provide much 
information about the effects of strong correlation of many levels.
This model was first studied by Carroll and Hioe \cite{CH86}, and
the exact transition probabilites have been obtained. 
The formulas for the probabilities show some varieties of expressions according to the 
relation between the slopes of diabatic states.
The generalized arbitrary $N$-level system for this model is called 
the bow-tie model \cite{ON97}, and the transition probabilities and 
characteristic mechanisms of transitions are discussed \cite{DO00}. 
Thus this model is quite convenient for comparison of the transition 
probabilities in the pure quantum case and the dissipative case.
We should also note that there exist some proposals for physical realization 
of this model by using optical systems \cite{CH86,HA97}.

\subsection{Analysis of Transition Probabilities}
We here enumerate exact transition probabilities in the three-level
system in the presence of dissipation. 
The Hamiltonian we consider is written for the diabatic bases:
\begin{eqnarray}
{\cal H} (t) &=& {\cal H}_0 (t) + \sum_{k}\xi_{k} (t) X_{k} \\
{\cal H}_0 (t) &=& \left(
\begin{array}{ccc}
0 & \Gamma_1 & \Gamma_2 \\
\Gamma_1 & v_1 t  & 0 \\
\Gamma_2 & 0 & v_2 t
\end{array}
\right) 
\qquad
{X}_{k} = \left(
\begin{array}{ccc}
a_{1}^{(k)} & 0 & 0 \\
0 & a_{2}^{(k)}  & 0 \\
0 & 0 & a_{3}^{(k)}
\end{array}
\right)
\end{eqnarray}
Here the $k$th white noise acts on the $k$th operator $X_{k}$. In this
case, the matrix element of the master equation (\ref{maseq}) is written
in the following form,
\begin{eqnarray}
\frac{\partial \rho_{\ell , m} (t)}{\partial t}
&=& -i \left( {H_0}(t)_{\ell , k }\rho_{k,m} (t) - 
              \rho_{\ell,k} (t){H_0}(t)_{k,m } \right)
- \bar{\gamma}_{\ell , m} \rho_{\ell , m} (t)  , \label{master3}
\end{eqnarray}
where $\bar{\gamma}_{\ell , m}$ is 
written using the matrix elements of the 
operator $X_k$ and the amplitude of the noise $\gamma_k$ as,
\begin{eqnarray}
\bar{\gamma}_{\ell , m} &=& \sum_{k}\frac{\gamma_{k}}{2}
\left( a_{\ell}^{(k)} - a_{m}^{(k)} \right)^2 .
\end{eqnarray}
Thus the differential equations for all matrix elements 
are given by,
\begin{eqnarray}
\dot{\rho_{00}}(t) &=& -i \left\{ \Gamma_1 (\rho_{10} -\rho_{01}) 
+\Gamma_2 (\rho_{20} - \rho_{02})\right\}  , \\
\dot{\rho_{11}}(t) &=& -i \Gamma_1 (\rho_{01} -\rho_{10}) , \\
\dot{\rho_{22}}(t) &=& -i \Gamma_2 (\rho_{02} -\rho_{10}) ,  \\
\dot{\rho_{01}}(t) &=& -i \left\{  \Gamma_1 ( \rho_{11} - \rho_{00} )
+\Gamma_2 \rho_{21} - v_1 t \rho_{01} \right\} -\bar{\gamma}_{01} \rho_{01} , \\
\dot{\rho_{02}}(t) &=& -i \left\{  \Gamma_2 ( \rho_{22} - \rho_{00} )
+\Gamma_1 \rho_{12} - v_2 t \rho_{02} \right\} -\bar{\gamma}_{02} \rho_{02} , \\
\dot{\rho_{12}}(t) &=& -i \left\{  \Gamma_1 \rho_{02} - \Gamma_2 \rho_{10}
+(v_1 - v_2 )t \, \rho_{12}  \right\} -\bar{\gamma}_{12} \rho_{12} .\\
\end{eqnarray}
Here we confine ourselves to the 
SD limit $\bar{\gamma}_{k,\ell }\rightarrow \infty$, and 
the initial condition $\rho_{01} (t_0) =\rho_{02} (t_0 )=\rho_{12}(t_0) =0$.
In the same manner as in the two-level case, $\rho_{12}(t)$ is
approximated, expanding by partial integrals,
\begin{eqnarray}
\rho_{12}(t) &=& \exp\left(-i\frac{(v_1 - v_2) t^{2}}{2} - \bar{\gamma}_{12}t \right)
\int_{t_0}^{t}\, du \,
\exp\left(i\frac{(v_1 - v_2) u^{2}}{2} + \bar{\gamma}_{12}u \right)
\left[ i\Gamma_{2} \rho_{10} (u) - i\Gamma_1\rho_{02}(u) \right]
 \nonumber \\
&\sim & \frac{i\Gamma_2 \rho_{10} (t)-i\Gamma_1 \rho_{02}(t) }{i(v_1 - v_2) t +\bar{\gamma}_{12}} . \label{r12}
\end{eqnarray}
$\rho_{02}(t)$ and $\rho_{01} (t)$ are written as,
\begin{eqnarray}
\rho_{02}(t) &=& \exp\left(i\frac{v_2 t^{2}}{2} - \bar{\gamma}_{02}t \right)
\int_{t_0}^{t}\, du \,
\exp\left(-i\frac{v_2 u^{2}}{2} + \bar{\gamma}_{02}u \right)
\left\{ -i\Gamma_{2} \left( \rho_{22} (u) - \rho_{00}(u) \right) - 
i\Gamma_{1} \rho_{12} (u)\right\}   \nonumber \\
&\sim & \frac{-i\Gamma_{2} \left( \rho_{22} (t) - \rho_{00} (t) \right) -i\Gamma_1 \rho_{12} (t)}{-i v_2 t + \bar{\gamma}_{02}} , \label{r02}\\
\rho_{01}(t) &=& \exp\left(i\frac{v_1 t^{2}}{2} - \bar{\gamma}_{01}t \right)
\int_{t_0}^{t}\, du \,
\exp\left(-i\frac{v_1 u^{2}}{2} + \bar{\gamma}_{01}u \right)
\left\{ -i\Gamma_{1} \left( \rho_{11} (u) - \rho_{00}(u) \right) - 
i\Gamma_{2} \rho_{21} (u)\right\}  \nonumber \\
&\sim &  \frac{-i\Gamma_{1} \left( \rho_{11} (t) - \rho_{00} (t) \right) -i\Gamma_2 \rho_{21} (t)}{-i v_1 t + \bar{\gamma}_{01}}  \label{r01}
\end{eqnarray}
The terms $\rho_{12} (t)$ and $\rho_{21} (t)$ in (\ref{r02}) and (\ref{r01}) 
are negligible because of Eq.(\ref{r12}).
Therefore we can approximate $(\ref{r02})$ and
$(\ref{r01})$ as,
\begin{eqnarray}
\rho_{02}(t) 
&=& \left(\frac{\Gamma_2}{v_2}\right) \left( \frac{1 }{t + i\frac{\bar{\gamma}_{02}}{v_2}} \right)
\left( \rho_{22}(t) - \rho_{00} (t) \right)  , \\
\rho_{01}(t)&=& \left(\frac{\Gamma_1}{v_1}\right) \left( \frac{1 }{t + i\frac{\bar{\gamma}_{01}}{v_1}} \right)
\left( \rho_{11}(t) - \rho_{00} (t) \right)  .
\end{eqnarray}
Thus we arrive at the effective master equations 
defining $c_1 (t) = \rho_{11}  (t) - \rho_{00} (t)
$ and $c_2  (t) = \rho_{22}(t) - \rho_{00} (t)$ :
\begin{eqnarray}
\dot{c}_{1} &=& \frac{-2i\Gamma_{1}^{2}  }{v_1} 
\left( \frac{1}{t+ i \bar{\gamma}_{01}/v_1} - \frac{1}{t-i\bar{\gamma}_{01}/v_1} \right)
c_{1} -\frac{i\Gamma_{2}^{2}  }{v_2} 
\left( \frac{1}{t+ i \bar{\gamma}_{02}/v_2} - \frac{1}{t-i\bar{\gamma}_{02}/v_2} \right)
c_{2} \label{ef1}\\
\dot{c}_{2} &=& \frac{-i\Gamma_{1}^{2}  }{v_1} 
\left( \frac{1}{t+ i \bar{\gamma}_{01}/v_1} - \frac{1}{t-i\bar{\gamma}_{01}/v_1} \right)
c_{1} -\frac{2i\Gamma_{2}^{2}  }{v_2} 
\left( \frac{1}{t+ i \bar{\gamma}_{02}/v_2} - \frac{1}{t-i\bar{\gamma}_{02}/v_2} \right)
c_{2}  \label{ef2}.
\end{eqnarray}
All matrix elements $\rho_{k, \ell} (t)$ are obtained if 
$\rho_{00}(t)$, $\rho_{11}(t)$, and $\rho_{22}(t)$ are 
calculated as solutions of these effective master
equation. 

Now let us solve these differential equations. 
We start with the parameters which satisfy the special relation,
\begin{eqnarray}
\frac{ \bar{\gamma}_{01}}{| v_1 | } = \frac{ \bar{\gamma}_{02}}{|v_2 |} = \alpha . \label{rel}
\end{eqnarray}
Here $\alpha $ is always positive since $\bar{\gamma}_{01} > 0$ and $\bar{\gamma}_{02} > 0$.
In this special case, the equations are simplified in the form,
\begin{eqnarray}
\frac{d}{d\, t}\left(
\begin{array}{c}
c_1 (t)
\\
c_2 (t)
 \end{array}
\right)
&=& - i \left( \frac{1}{t+i \alpha} - \frac{1}{t-i \alpha}\right)  
M
\left(
\begin{array}{c}
c_1 (t)
\\
c_2 (t)
 \end{array}
\right) , \label{ef_master3}
\end{eqnarray}
where the matrix $M$ is given by
\begin{eqnarray}
M &=& \left(
\begin{array}{cc}
2a& b\\
a
& 2b
\end{array}
\right) , \qquad \qquad 
a=  \frac{\Gamma_{1}^{2}}{| v_1 |}  \quad  {\rm and} \quad b  =
\frac{\Gamma_{2}^{2}}{| v_2 |} .
\end{eqnarray}
The right hand side of eq.(\ref{ef_master3}) has the explicit time 
dependence only in the prefacer.
Therefore by diagonalizing the matrix $M$, we can obtain the scattering matrix
which connects $c(-\infty)$ with $c(\infty)$.
The matrix $M$ has these eigenvalues $\lambda_{\pm}$
\begin{eqnarray}
\lambda_{\pm} = a +b \pm \sqrt{a^2 +b^2 - ab} .
\end{eqnarray}
Using $a$, $b$, and these eigenvalues $\lambda_{\pm}$, 
the final state and initial state are connected using 
the scattering matrix $S$ as,
\begin{eqnarray}
\left(
\begin{array}{c}
c_1 (\infty )
\\
c_2 (\infty )
 \end{array}
\right)
&=& S 
\left(
\begin{array}{c}
c_1 (-\infty )
\\
c_2 (-\infty )
 \end{array}
\right) , \\
S_{1,1} &=& \frac{1}{2\sqrt{a^2 + b^2 -ab}}=
\left[ (a-b)(e^{-2\pi\lambda_{+}} - e^{-2\pi\lambda_{-}})
+ \sqrt{a^2 + b^2 -ab } (e^{-2\pi\lambda_{+}} + e^{-2\pi\lambda_{-}} )
\right] , \label{s11} \\
S_{1,2} &=& \frac{b}{2\sqrt{a^2 + b^2 -ab}} ( e^{-2\pi\lambda_{+}} -
e^{-2\pi\lambda_{-}} ) ,  \\
S_{2,1} &=& \frac{a}{2\sqrt{a^2 + b^2 -ab}} ( e^{-2\pi\lambda_{+}} -
e^{-2\pi\lambda_{-}})  ,  \\
S_{2,2} &=& \frac{1}{2\sqrt{a^2 + b^2 -ab}}
\left[ (a-b)(-e^{-2\pi\lambda_{+}} + e^{-2\pi\lambda_{-}})
+ \sqrt{a^2 + b^2 -ab } (e^{-2\pi\lambda_{+}} + e^{-2\pi\lambda_{-}})
\right] \label{s22}.
\end{eqnarray}
Here $\alpha $ does not appear because it only gives the singular point
in the Cauchy's integral to yield (\ref{s11})-(\ref{s22}),  
which is the same situation as in the 
two-level system. Therefore the scattering matrix does not depend on 
the concrete values of dissipation strength, $\bar{\gamma}_{01}$ and $\bar{\gamma}_{02}$
as far as those are large and the relation (\ref{rel}) is satisfied. 
We now obtain the probabilities for various initial states.
\begin{eqnarray}
P^{-+}_{\rm SD} (0\rightarrow 0)   &=&\frac{1}{3} 
+ \frac{a+b}{6\sqrt{a^2 + b^2 -ab}}\, (e^{-2\pi\lambda_{+}} - e^{-2\pi\lambda_{-}} ) + \frac{e^{-2\pi\lambda_{+}} + e^{-2\pi\lambda_{-}}}{3} \label{p00}\\
P^{-+}_{\rm SD} (0\rightarrow 1) = P^{-+}_{\rm SD} (1\rightarrow 0)&=&\frac{1}{3} 
+ \frac{-2a+b}{6\sqrt{a^2 + b^2 -ab}}\, (e^{-2\pi\lambda_{+}} - e^{-2\pi\lambda_{-}} ) - \frac{e^{-2\pi\lambda_{+}} + e^{-2\pi\lambda_{-}}}{6} \\
P^{-+}_{\rm SD} (0\rightarrow 2) = P^{-+}_{\rm SD} (2\rightarrow 0) &=&\frac{1}{3} 
+ \frac{a-2b}{6\sqrt{a^2 + b^2 -ab}}\, (e^{-2\pi\lambda_{+}} - e^{-2\pi\lambda_{-}} ) - \frac{e^{-2\pi\lambda_{+}} + e^{-2\pi\lambda_{-}}}{6} \\
P^{-+}_{\rm SD} (1\rightarrow 0) = P^{-+}_{\rm SD} (0\rightarrow 1) &=&\frac{1}{3} 
+ \frac{-2a+b}{6\sqrt{a^2 + b^2 -ab}}\, (e^{-2\pi\lambda_{+}} - e^{-2\pi\lambda_{-}} ) - \frac{e^{-2\pi\lambda_{+}} + e^{-2\pi\lambda_{-}}}{6} \\
P^{-+}_{\rm SD} (1\rightarrow 1) &=&\frac{1}{3} 
+ \frac{a-2b}{6\sqrt{a^2 + b^2 -ab}}\, (e^{-2\pi\lambda_{+}} - e^{-2\pi\lambda_{-}} ) + \frac{e^{-2\pi\lambda_{+}} + e^{-2\pi\lambda_{-}}}{3} \\
P^{-+}_{\rm SD} (1\rightarrow 2) = P^{-+}_{\rm SD} (2\rightarrow 1) &=&\frac{1}{3} 
+ \frac{a+b}{6\sqrt{a^2 + b^2 -ab}}\, (e^{-2\pi\lambda_{+}} - e^{-2\pi\lambda_{-}} ) - \frac{e^{-2\pi\lambda_{+}} + e^{-2\pi\lambda_{-}}}{6} \\
P^{-+}_{\rm SD} (2\rightarrow 2) &=&\frac{1}{3} 
+ \frac{-2a+b}{6\sqrt{a^2 + b^2 -ab}}\, (e^{-2\pi\lambda_{+}} - e^{-2\pi\lambda_{-}} ) - \frac{e^{-2\pi\lambda_{+}} + e^{-2\pi\lambda_{-}}}{3} \label{p22},
\end{eqnarray}

In the adiabatic limit $v_1 \rightarrow +0$ and $v_2 \rightarrow +0$, 
all probabilities converge
to $1/3$ due to the strong effect of dissipation. As shown numerically 
in the next section, these formulas are always valid even when the 
relation (\ref{rel}) is not satisfied. That is, the probabilities are little affected by the variation of the strength of dissipation $\bar{\gamma}_{01}, 
\bar{\gamma}_{012}$, and $\bar{\gamma}_{12}$. 

\subsection{Numerical Investigation}

We numerically integrate the equation (\ref{master3}), 
and compare with the asymptotically exact transition probabilities obtained above for various parameter values.
We write the slopes of the diabatic states $v_1$ and $v_2$ using parameter
$v$,
\begin{eqnarray}
v_1 &=& \alpha_1 v ,\\
v_2 &=& \alpha_2 v .
\end{eqnarray}  
Dimensionless parameters $\alpha_1$ and $\alpha_2$ gives the ratio of
$v_1$ to $v_2$ and determine the level structure. 
We consider three types of level structures, 
namley (a)$ \,\, \alpha_1 \cdot\alpha_2
<0$,  (b)$ \,\,  \alpha_1  > \alpha_2 > 0$, and 
(c)$ \,\,\alpha_2  > \alpha_1 > 0$.
In Fig.2, the transition probabilities $P^{-+}_{\rm SD} (0\rightarrow
j) \, (j=0,1,2)$ are shown for these cases. 
The probabilities are plotted as a function of the parameter $v$
for the parameters $\Gamma_{1}=0.1$, $\Gamma_{2}=0.2$, and various sets of 
$(\bar{\gamma}_{01},\bar{\gamma}_{02},\bar{\gamma}_{12})$. 
Here the sets of $(\alpha_1 , \alpha_2)$ are taken as $(1,-0.5)$ 
for Fig.2(a), $(1,0.5)$ for Fig.2(b), and $(0.5,1)$ for Fig.2(c), 
respectively. 
The lines are theoretical values for the SD limit $P^{-+}_{\rm SD}
(0\rightarrow j)$ given by (\ref{p00})-(\ref{p22}) and 
the probabilities for pure quantum case 
$P^{-+}_{\rm LZ} (0\rightarrow j)$ which are already obtained by 
Carroll and Hioe \cite{CH86}. The analytical solutions 
of probabilities in the pure quantum case are listed in
Table I.  
As can be seen in these figures, we can see good agreements 
between the numerical data and theories (\ref{p00})-(\ref{p22}). 
We find little dependence on the variety of 
$(\bar{\gamma}_{01},\bar{\gamma}_{02},\bar{\gamma}_{12})$, that is,
the formula (\ref{p00})-(\ref{p22}) are valid even if
the relation (\ref{rel}) is not satisfied as far as 
the dissipation is very strong. This was also confirmed for 
$P^{-+}_{\rm SD} (1\rightarrow j)$, and $P^{-+}_{\rm SD} (2\rightarrow j)$.
\begin{figure}
\begin{center}
\noindent
\epsfxsize=9.0cm \epsfysize=6.0cm
\epsfbox{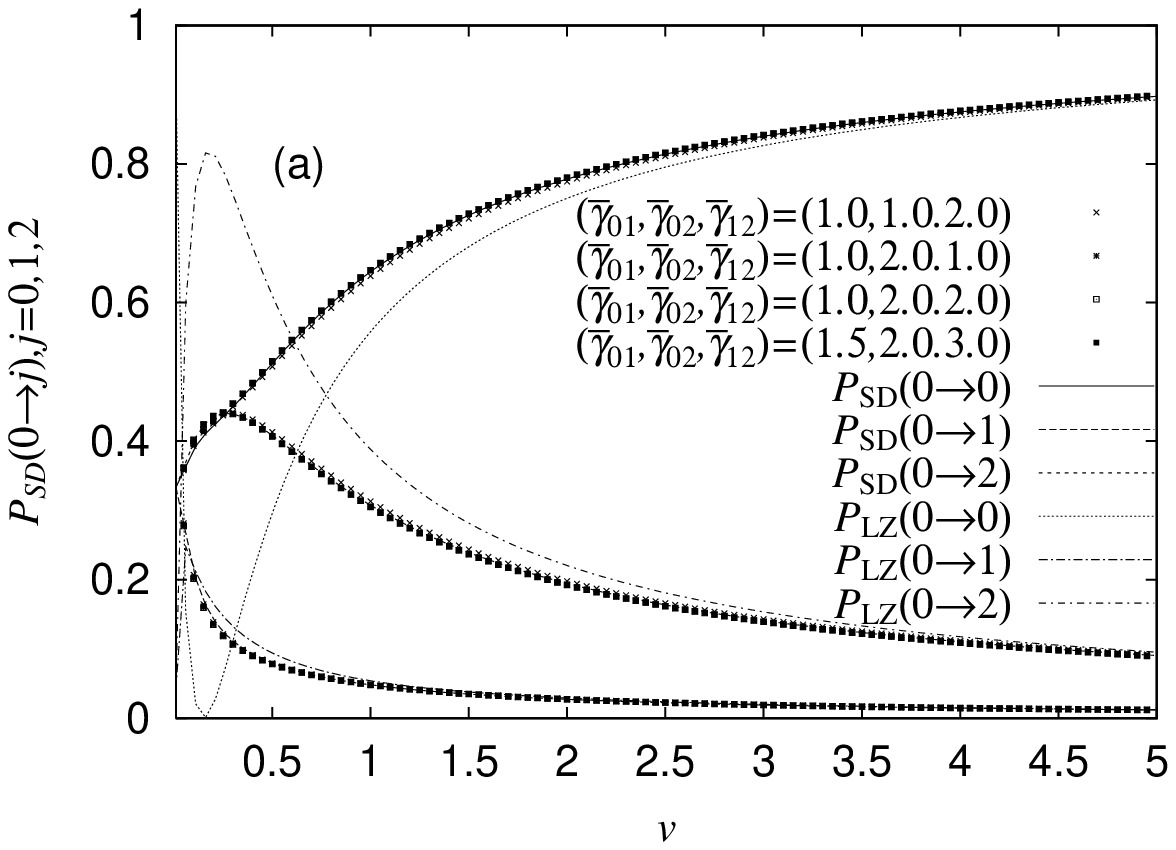}
\epsfxsize=9.0cm \epsfysize=6.0cm
\epsfbox{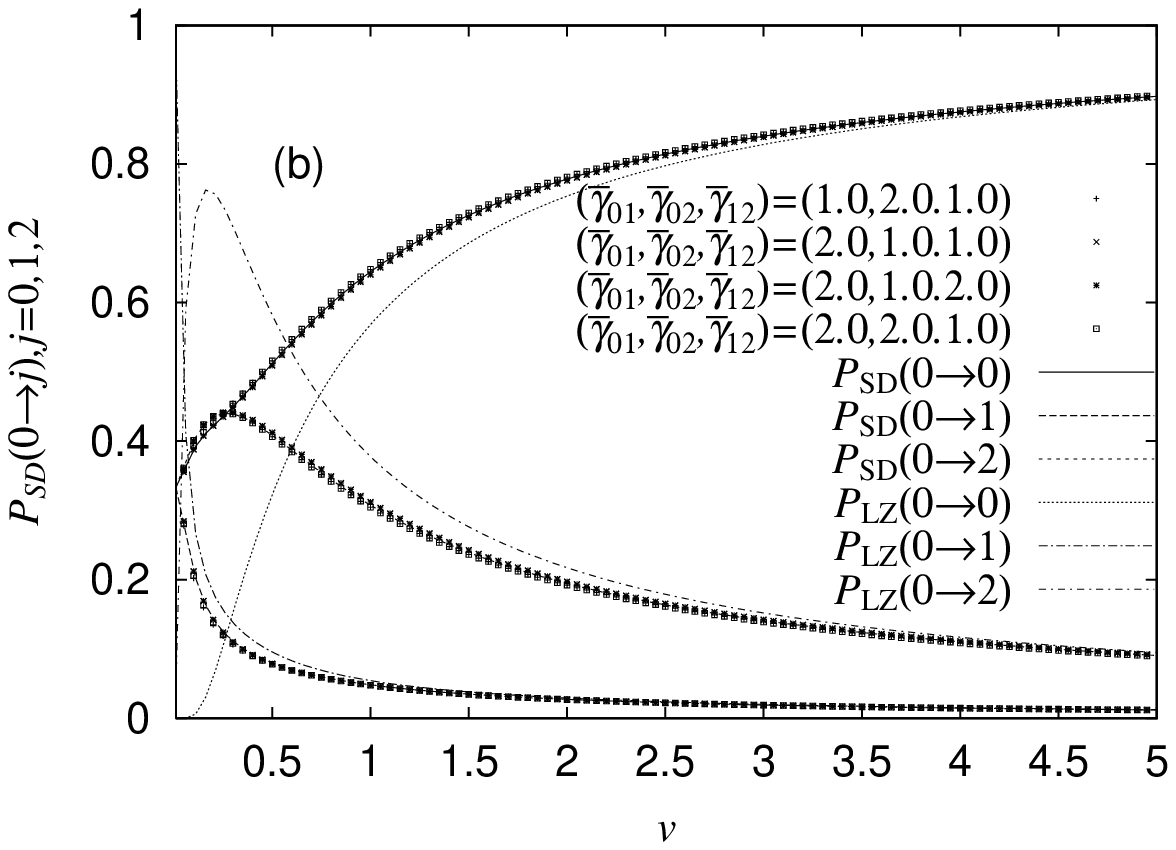} \\
\epsfxsize=9.0cm \epsfysize=6.0cm
\epsfbox{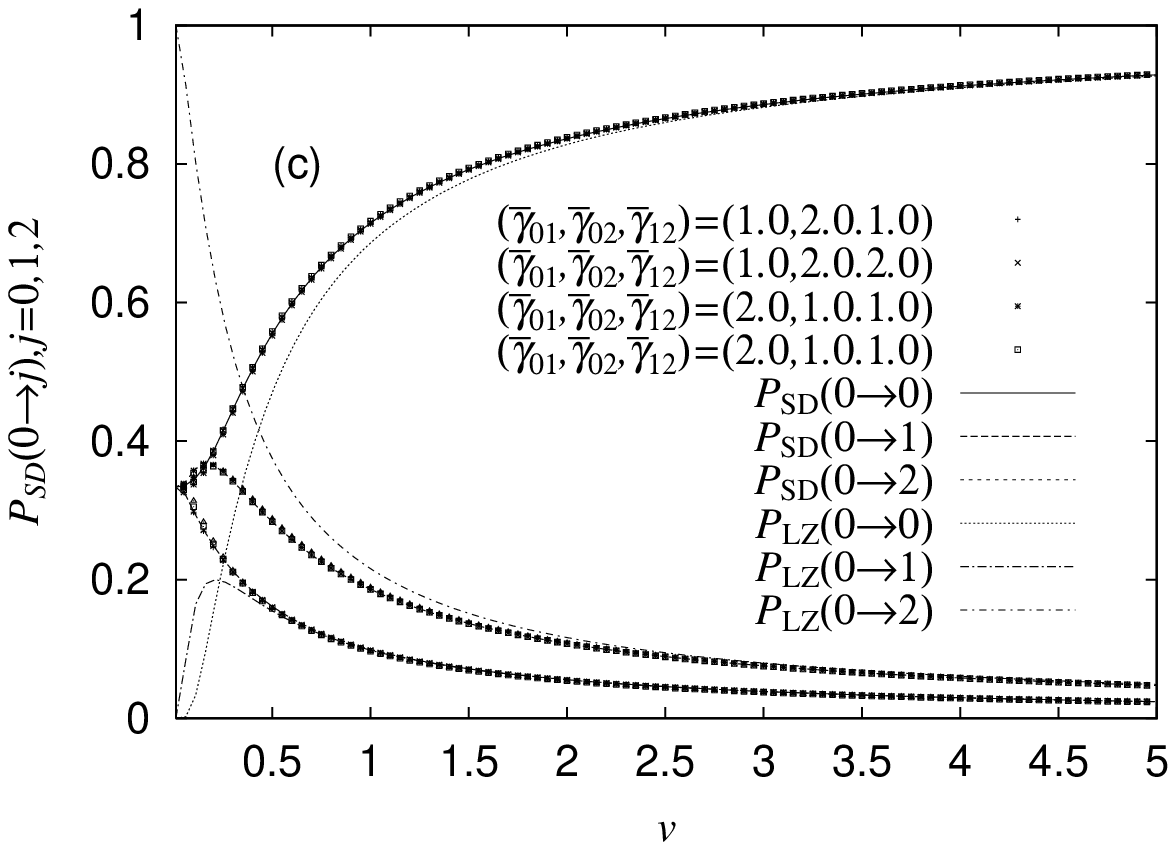}
\end{center}
\caption{Comparison the numerical calculation with theories for various
 sets of $\{\bar{\gamma}_{ij}\}$. Data are plotted as a function of $v$.
Points are numerical data,  and the lines of $P^{-+}_{\rm SD}$ 
and $P^{-+}_{\rm LZ}$ (see Table I) are theoretical values. 
Fig.1(a): $(\alpha_1 , \alpha_2) = (1,-0.5) $, Fig.2(b): $(1,0.5)$,
Fig.2(c): $(0.5,1)$. }
\end{figure}
As seen in Table I, the analytical expressions in pure quantum case 
$P^{-+}_{\rm LZ} (0\rightarrow j)$ show some variations according to 
relations of $v_1$ and $v_2$.
On the other hand, the probabilities in the dissipative case 
(\ref{p00})-(\ref{p22}) do not depend on such level structures.
For instance, in cases of Fig.2(a) and Fig.2 (b) 
where the sweeping velocities are
$(v_1 , v_2) = (v, -0.5v)$ and $(v_1 , v_2) = (v, 0.5v)$, respectively,  
each probability $P^{-+}_{\rm SD} (i\rightarrow j)$ 
for the both cases completely agrees with each other 
because formulae  
(\ref{p00})-(\ref{p22}) have the same values for different $(v_1, v_2)$
with the same absolute values. 
However in the pure quantum case, 
the probabilities $P^{-+}_{\rm LZ} (i\rightarrow j)$ are different 
between these cases as found in Table I.
This is a remarkable contrast between the dissipative case and the pure
quantum case.

In the slow sweeping $v \ll 1$, the deviation of $P^{-+}_{\rm SD}$ from
$P^{-+}_{\rm LZ}$ are large. In the fast sweeping region $v\gg 1$,
$P^{-+}_{\rm SD}$ asymptotically converges to the behavior 
of $P^{-+}_{\rm LZ}$.
This means that the system is little affected from dissipation for
fast sweeping because the time that the system stays at around 
avoided level crossing point is very short. This is the same 
behavior as found in the two-level system in \cite{K84}. 
\begin{table}[ht]
\noindent
\caption{The transition probabilities for quantum case. Here
 $P=\exp\left(-\pi \Gamma_{1}^{2} / v_1 \right)$ and $Q=\exp\left(-\pi \Gamma_{2}^{2} / v_2 \right)$. }
\begin{center}
\begin{tabular}{|c|c|c|c|c|c|c|c|}
%\hline
\quad 
& $P^{-+}_{C} (0\rightarrow 0) $
& $P^{-+}_{C} (0\rightarrow 1) $
& $P^{-+}_{C} (0\rightarrow 2) $
& $P^{-+}_{C} (1\rightarrow 1) $
& $P^{-+}_{C} (1\rightarrow 2) $
& $P^{-+}_{C} (2\rightarrow 2) $ \\
\hline
$v_1 \cdot v_2 < 0$ 
& $\left( 1-P-Q \right)^2$
& $\left( 1-P\right)\left( P + Q \right)$
& $\left( 1-Q\right)\left( P + Q \right)$
& $ P^2$
& $\left(1-P\right)\left( 1-Q \right)$
& $Q^2$ \\
\hline
$|v_1| > |v_2|,v_1 \cdot v_2 > 0$ 
& $ P^2 Q^2$
& $\left( 1-P\right)\left( 1+PQ \right)$
& $P\left( 1-Q\right)\left( 1+PQ \right)$
& $ P^2$
& $ P\left(1-P\right)\left( 1-Q \right)$
& $\left( 1-P+PQ \right)^2$ \\
\hline
$|v_2| > |v_1| , v_1 \cdot v_2 > 0$ 
& $ P^2 Q^2$
& $Q\left( 1-P\right)\left( 1+PQ \right)$
& $\left(1-Q \right) \left( 1+PQ \right)$
& $\left( 1-Q+PQ \right)^2$
& $ Q\left(1-P\right)\left( 1-Q \right)$
& $ Q^2$  \\
\end{tabular}
\end{center}
\end{table}

\section{Summary}
In two and three-level system, we derived the effective master equations
which well describe time evolution of the system in the SD limit.
Thereby we obtained analytical transition probabilities. 
The effective master equation approach is quite useful
because the differential equation of system's variable becomes very simple.
This approach will be applicable in the other systems 
whose exact transition probabilities can be analytically enumerated in the 
pure quantum case. 

In the two-level system, we consider the two cases where 
the external field is swept from large negative field and from zero field.
Both situations are easily realized in real experiments.
We hope that the exact relation (\ref{ene1}) and (\ref{ene2}) is confirmed 
in real experiments using classical optical system \cite{RKLK81} and 
Cooper pair \cite{NEC}, and so on.

The transition time in the two-level systems has been discussed 
in the literatures \cite{K84,MBGS89,V99} when the external field is reversed
from large negative field. 
According to Vitanov's definition \cite{V99},
the transition times $t^{\rm tr}$ is written as,
\begin{eqnarray}
t^{\rm tr} \equiv \frac{\rho_{22} (\infty)}{\rho_{22}' (0)} &=&
\frac{c_1 (\infty )}{c_1' (0)} \label{v_def}, 
\end{eqnarray}
under the initial condition of $\rho_{11} (-\infty ) =1$. Vitanov derived
the exact expression of transition time $t^{\rm tr}_{\rm LZ}$ 
in the pure quantum case as follows, 
\begin{eqnarray}
t^{\rm tr}_{\rm LZ} &=& {\sqrt{1 - e^{-\pi \Gamma^2 / v}}\over (\Gamma/v )
\cos (\chi )} ,\label{vitanov} \\
\chi &=& \frac{\pi}{4} 
+ \arg {\bf \Gamma} \left( {1\over2} - {i\Gamma^2\over 4 v}\right)
- \arg {\bf \Gamma} \left(         1 - {i\Gamma^2\over 4 v}\right) ,
\end{eqnarray}
where ${\bf \Gamma} (x)$ is the Gamma function. Eq.(\ref{vitanov}) 
converges to 
\begin{eqnarray}
t^{\rm tr}_{\rm LZ} \rightarrow 
\left\{
\begin{array}{cc}
\sqrt{2\pi\over v} & \cdots {\Gamma^2\over v} \ll 1 \\
2{\Gamma\over v}   & \cdots {\Gamma^2\over v} \gg 1 \\
\end{array}
\right. \label{limit} .
\end{eqnarray}
This asymptotic behavior agrees with the one in the earlier studies 
\cite{K84,MBGS89}.
On the other hand, the transition time of dissipative case $t^{\rm
tr}_{\rm SD}$ is readily written for the definition ({\ref{v_def}});
\begin{eqnarray}
t^{\rm tr}_{\rm SD} = {\Gamma^2\over \gamma}
\exp\left(-{\pi\Gamma^2\over 2 v} \right) .
\end{eqnarray}
This means that the transition time decreases as $O(1/\gamma)$ 
with the increase of $\gamma$. 
In the three-level system, the analytical expressions (\ref{p00})-(\ref{p22})
are the solutions under the condition (\ref{rel}). However these
solutions are valid beyond the condition (\ref{rel}) as shown numerical 
calculation. According to Carroll and Hioe's analytical solution (Table
I), the pure quantum transition probabilities show some variations 
depending on the level structures. 
However, the probabilities in the dissipative case do not show such dependences.
It is interesting to confirm experimentally this thermal effect because 
the three-level system we consider can be realized 
experimentally \cite{CH86,HA97}. 

Eq.(\ref{maseq}) is derived under the condition where
the noise affects on the system only through the diabatic states. 
In case of the off-diagonal coupling, i.e., $X_{\ell} = \sigma^{x}$, 
we also derived the similar master equation. 
In this case, we can easily show that in the strong damping limit, 
all the transition probabilities become uniform regardless of 
initial condition \cite{N01,k_yoko}.
Thus the transition probability is affected by the coupling form with
the thermal environment. 
Therefore it is also interesting to consider the transition probabilities for 
various coupling forms with finite $\gamma$.

\section*{Acknowledgment}
One of the authors (K.S.) would like to thank Professor Seiji Miyashita for
valuable comments. The present work is partially supported by the 
Grand-in-Aid for Scientific Research from Ministry of Education,
Science, Sports and Culture of Japan.

\section*{Appendix A}
The same type of master equation as Eq. (\ref{MbE}) is derived
in the case where 
the system of interest couples with the phonon bath
through the diabatic states as follows \cite{CL83},
\begin{eqnarray}
{\cal H}_{tot} = {\cal H} (t) + \lambda \sum_{\ell} X_{\ell} \sum_{\omega}
 \gamma_{\alpha} \left( {b_{\omega}^{(\ell )}}^{\dagger}  
+ b_{\omega}^{(\ell )} \right)  + \sum_{\ell , \omega} 
\omega {b_{\omega}^{(\ell )}}^{\dagger} 
b_{\omega}^{(\ell )} \label{tsys},
\end{eqnarray}
where ${\cal H} (t) $ denotes the system Hamiltonian of interest,
and $X_{\ell}$ is the $\ell$th operator which interacts with 
the phonon system. We assume the computability between the coupling
operators $X_{\ell}$, i.e, $\left[ X_{\ell} , X_{\ell '}\right] = 0$.
The operator ${b_{\omega}^{(\ell )} }^{\dagger} $ 
and $b_{\omega}^{(\ell )} $ is the creation and annihilation phonon operator 
which interacts with the system through the $\ell$th coupling operator $ X_{\ell}$.

We use the projection operator technique to trace out 
the reservoir's degree of freedom \cite{NZM}and 
assume that the correlation between reservoir's variables is
short-lived. Then we obtain the master 
equation for the system in the second order of coupling strength
$\lambda$ \cite{WH65,QME};
\begin{equation}
\frac{\partial}{\partial t} \rho(t) = 
\frac{1}{i\hbar}\left[ {\cal H},\rho(t)\right] 
- \lambda^{2}\sum_{\ell}\Gamma_{\ell}\rho(t),
\label{ME}
\end{equation}
where $\Gamma_{\ell}\rho(t)$ is given by
\begin{eqnarray}
\Gamma_{\ell}\rho(t) &=& \frac{1}{\hbar^{2}}
\int_{0}^{\infty}\, dt'\int_{-\infty }^{\infty }\, d\omega \,
e^{i\omega t'} \Phi_{\ell}(\omega)\left\{ X_{\ell}X_{\ell}
(-t')\rho(t) \right. \nonumber \\
& & \left. \mbox{}- e^{\beta\hbar\omega}X_{\ell}\rho(t)X_{\ell}(-t') 
+ e^{\beta\hbar\omega}\rho(t)X_{\ell}(-t')X_{\ell} 
- X_{\ell}(-t')\rho(t)X_{\ell}
\right\} , 
\label{DST}
\end{eqnarray}
Here $X_{\ell}(-t')$ means the Heisenberg operator at time $-t'$, 
\begin{equation}
X_{\ell}(-t') = \exp_{\leftarrow}\left(-\frac{i}{\hbar}\int^{0}_{-t'}\, 
du \, H(u) \right) X  
\exp_{\rightarrow}\left(\frac{i}{\hbar}\int^{0}_{-t'}\, du \, H(u) \right)
\end{equation}
In case of phonon reservoir described in (\ref{tsys}), $\Phi (\omega )$
is given by, 
\begin{eqnarray}
\Phi_{\ell} (\omega ) &=& \hbar\frac{I_{\ell}(\omega)-I_{\ell}(-\omega)}
{e^{\beta\hbar\omega }-1} ,
\label{phiomega}
\end{eqnarray}
where $\beta$ is an inverse temperature $1/T$.
$I_{\ell}(\omega)$ is called the spectral density.

We restrict ourselves to the case of the Ohmic spectrum,
\begin{eqnarray}
I_{\ell}(\omega) = I_{\ell} \omega , 
\end{eqnarray}
and high temperature,
\begin{eqnarray}
 T \gg 1 .
\end{eqnarray}
In this case, by using the fact,
\begin{eqnarray}
\Phi_{\ell} (\omega ) &\rightarrow& \hbar I_{\ell} T ,
\label{phiomega}
\end{eqnarray}
the master equation (\ref{ME}) is reduced to the simple form,
\begin{eqnarray}
\frac{\partial}{\partial t} \rho(t) = 
\frac{1}{i\hbar}\left[ {\cal H},\rho(t)\right] 
- \lambda^{2}T \sum_{\ell} I_{\ell} 
\left[ X_{\ell}, \left[ X_{\ell} , \rho (t) \right]\right] .
\label{MbE}
\end{eqnarray}
Although the equation (\ref{DST}) derived by the projection operator 
approach is an approximation because higher order
terms of $\lambda$ is neglected and the fast relaxation of reservoir is
assumed to make the equation Markovian, it can well describe 
features of evolution of the system especially in case of high temperature.

\section*{Appendix B}
In this Appendix, the transition probabilities (\ref{lz0+}) are
derived. We start with the Hamiltonian, 
\begin{eqnarray}
{\cal H}(t) &=& \frac{\hbar}{2}\left( \Gamma \sigma^x - v t \sigma^z 
\right)  .
\end{eqnarray}
We consider the Schr\"{o}dinger equation.  
\begin{eqnarray}
i\hbar \frac{\partial }{\partial t} \Psi (t)
&=& \frac{\hbar}{2}\left( \Gamma \sigma^x - v t \sigma^z 
\right) \Psi (t) \label{shre} .
\end{eqnarray}
This equation (\ref{shre}) is concretely written defining the
component like $\Psi (t) = (x_1 (t) , x_2 (t) )^{\dagger}$,
\begin{eqnarray}
i \dot{x}_{1} (t) &=& \frac{vt}{2} x_1 (t) + \frac{\Gamma }{2} x_2 (t) 
\label{eq1}\\
i \dot{x}_{2} (t) &=& \frac{\Gamma }{2} x_1 (t) - \frac{vt}{2} x_2 (t) 
\label{eq2} .
\end{eqnarray}
Now we transform the variables to the following ones,
\begin{eqnarray}
\tau &:=& t^2 \\
y_{1} (\tau ) &:=& \frac{x_1 (t)}{t} \\
y_{2} (\tau ) &:=& x_2 (t) ,
\end{eqnarray}
then we obtain the equation, 
\begin{eqnarray}
 2i\tau \frac{d}{d\tau} y_1 + (i - \frac{v\tau}{2}) y_1 
- \frac{\Gamma }{2} y_2 &=& 0 \label{eq11}\\
 2 i \frac{d}{d\tau} y_2 + \frac{v}{2} y_2 
- \frac{\Gamma }{2} y_1 &=& 0 \label{eq12} .
\end{eqnarray}
Here we used the relation $d/dt = 2t \, d/d\tau$. By using the Laplace transformation 
\cite{M32} defined as,
\begin{eqnarray}
y_{k} (\tau ) = \int_{C_{\xi}}\, d\xi \, \tilde{x}_{k} (\xi ) e^{\xi \tau} ,
\end{eqnarray}
we obtain the integral representation for $x_1 (t) $ and $x_2 (t)$ after
straight-forward calculation,
\begin{eqnarray}
x_1 (t) &=& At \int_{C_\xi}\, 
d\xi \, 
\left( \xi + i \frac{v}{4} \right)^{-\frac{1}{2} + i \frac{\Gamma^2}{8v} }
\,
\left( \xi - i \frac{v}{4} \right)^{-i \frac{\Gamma^2}{8v} } \,
e^{\xi t^2} \\
x_2 (t) &=& A \left( -i\frac{\Gamma}{4}\right) \int_{C_\xi}\, 
d\xi \, 
\left( \xi + i \frac{v}{4} \right)^{-\frac{1}{2} + i \frac{\Gamma^2}{8v} }
\,
\left( \xi - i \frac{v}{4} \right)^{-1 -i \frac{\Gamma^2}{8v} } \,
e^{\xi t^2} .
\end{eqnarray}
Here the integral counter must satisfies the following condition,
\begin{eqnarray}
 \left[ \left( 2 i \xi - \frac{v}{2} \right)\tilde{x}_{1} (\xi )
e^{\xi\tau}  \right]_{C_\xi} = 0 .
\end{eqnarray}
For the variables $\xi = |\xi|e^{i \phi}, t = |t| e^{i\theta}$,
we choose the counter with the condition, 
\begin{eqnarray}
\phi + 2\theta &=& 3\pi , \label{cont}
\end{eqnarray}
for large $|\xi|$ noting the relation $\xi t = |\xi||t|e^{i(\phi + 2\theta)}$.
When the initial time is $t=\infty \cdot e^{i\pi}$ and 
the final time is $t=\infty \cdot e^{i 0}$, 
the phase of $\phi$ varies from $\pi$ to $3\pi$ from the relation (\ref{cont}). 
We now consider the initial condition as,
\begin{eqnarray}
x_1 (-\infty) = 1. \label{init_con}
\end{eqnarray}
This condition is realized in the contour which encircles 
the singular point $\xi = -i {v\over4}$ and choosing the constant $A$ as,
\begin{eqnarray}
A = e^{{\pi \Gamma^2\over 16v}}
/\int_{\infty}^{(0+)} \, du \, (-u)^{-1/2 +i\Gamma^2/8v} e^{-u}.
\end{eqnarray}
Thus the wave function at $t= \infty$ is calculated by using 
analytical continuation following the condition (\ref{cont}), 
i.e. $\pi\to\phi\to 3\pi$ \cite{M32}.  
Thus revival probability is calculated as,
\begin{eqnarray}
x_{1} ( |t|e^{i0} ) = x_{1} (|t|e^{i\pi}) \exp\left( -{\pi \Gamma^2\over 4v}\right)
\qquad (|t| \rightarrow \infty)
\end{eqnarray}
This means nothing but the relation of Landau-Zener transition.

Next we consider the case where the external field is swept from zero value
\cite{CH86,V96}, that is, $\theta = 0$ and $+0 \to |t| \to \infty$.
In order to derive the survival probability in this case, 
we first note the time symmetry that the probability is the same value as 
the one obtained when the initial time is taken as $t= -\infty$ 
(the external value $= -\infty$) and the final time is $t= -0 $ 
(the external value $= -0$). 
We consider the latter case $(-\infty\to t \to -0)$ because
we can use the same contour as the previous case (\ref{cont}) which satisfies
the initial condition (\ref{init_con}).
We can readily write the integral representation of the $x_1 (-0)$ and 
obtain the transition probability $P_{C}^{0+}$ :
\begin{eqnarray}
x_1 (-0) &=& A|t|\int_{\infty}^{(0+)} (-x)^{-1/2 +i\Gamma^2/8v}
\left( -x - i {v\over2}\right)^{-i\Gamma^2/8v} \, e^{-x -iv/4}|t|^2
\nonumber \\
&=& A \left( 1 + e^{\pi \Gamma^2 /4v}\right)
\int_{0}^{\infty} \, du u^{-1/2} e^{-u} \nonumber 
= \sqrt{ \frac{ 1 + e^{\pi \Gamma^2 /4v} }{2}} .
\end{eqnarray}

\begin{thebibliography}{notitle}
\bibitem{note} This approximation should be justified under the condition 
that the transition time in the pure quantum case is much larger than 
the dissipation time scale $1/\gamma$. See (\ref{vitanov}) 
for the tunneling time in the pure quantum case.
\bibitem{k_yoko} Y. Kayanuma, J. Phys. Soc. Jpn {\bf 54} 2037 (1985). 
This paper also treats  
the case of the off-diagonal coupling, and shows the nonuniform transition
probability. We should note that this result is derived in the different 
limiting case from ours. See also the reference \cite{N01}
\end{thebibliography}
\end{document}